\documentclass[12pt,onecolumn]{emulateapj}

\shorttitle{Relativistic synchrotron equipartition}
\shortauthors{Barniol Duran, Nakar \& Piran}

\usepackage{epsf}    
\usepackage{epsfig}
\usepackage{amsmath}
\usepackage{color} 
\usepackage{ulem}

\newcommand{\gae}{\lower 2pt \hbox{$\, \buildrel {\scriptstyle >}\over {\scriptstyle
\sim}\,$}}
\newcommand{\lae}{\lower 2pt \hbox{$\, \buildrel {\scriptstyle <}\over {\scriptstyle
\sim}\,$}}

\begin{document}

\title{Radius constraints and minimal equipartition energy of
  relativistically moving synchrotron sources}
\author{Rodolfo Barniol Duran\altaffilmark{1a}, Ehud Nakar\altaffilmark{2b} \& Tsvi Piran\altaffilmark{1c}}
\altaffiltext{1}{Racah Institute for Physics, The Hebrew University, Jerusalem, 91904, Israel}
\altaffiltext{2}{The Raymond and Berverly Sackler School of Physics and Astronomy,
Tel Aviv University, 69978 Tel Aviv, Israel} 
\email{(a) rbarniol@phys.huji.ac.il; (b) udini@wise.tau.ac.il; (c) tsvi.piran@mail.huji.ac.il}

\begin{abstract}
A measurement of the synchrotron self-absorption flux and frequency
provides tight constraints on the physical size of the source and a
robust lower limit on its energy. This lower limit is also a good
estimate of the magnetic field and electrons' energy, if
the two components are at equipartition. This well-known method was
used for decades to study numerous astrophysical 
sources moving at non-relativistic (Newtonian) speeds. 
Here we generalize the  Newtonian equipartition theory to sources moving at
relativistic speeds including the effect of deviation
from spherical symmetry expected in such sources. 
Like in the Newtonian case, minimization of the energy provides an excellent
estimate of the emission radius and yields a useful lower limit on the energy. 
We find that the application of the Newtonian formalism to a relativistic source 
would yield a smaller emission radius, and would generally yield a larger 
lower limit on the energy (within the observed region).
For sources where the Synchrotron-self-Compton component
can be identified, the minimization of the total energy is not necessary
and we present an unambiguous solution for the parameters of the system.
\end{abstract}

\section{Introduction}

The equipartition method (Pacholczyk 1970; Scott \& Readhead 1977;
Chevalier 1998) has been extensively applied to radio observations of
sources moving at non-relativistic speeds (we
refer to them as ``Newtonian sources''). In particular, it has been
applied to radio emission from
supernovae (e.g., Shklovskii 1985, Slysh 1990, Chevalier 1998,
Kulkarni et al. 1998, Li \& Chevalier 1999, Chevalier \& Fransson 2006, 
Soderberg et al. 2010a). The method relies on the fact that both the electron
and magnetic field energy of a system, which emits self-absorbed
synchrotron photons, depend sensitively on the source
size. This allows for a robust determination of the size and of the minimal
total energy needed to produce the observed emission. If the 
electron and magnetic field energies are close to equipartition then this
lower limit is also a good estimate of their true energy.
The strength of these arguments is that they
are insensitive to the origin of the conditions within the emitting source
and, as such, the results are independent of the details of the model.

The method depends only on the assumption of self-absorbed synchrotron emission. In
Newtonian sources it characterizes the emitting region with four unknowns.
Three are microphysical: the number of electrons\footnote{The calculations
here are insensitive to the charge sign
of the radiating particles, so if positrons are present then anywhere we refer
to electrons we actually refer to pairs.} that radiate in the observed
frequency, their Lorentz Factor (LF) and the
magnetic field. The macrophysical unknowns are the area and volume of the
emitting region, which is assumed to be spherical and thus are both expressed
by the fourth unknown: the source radius, $R$.
An observed synchrotron spectrum, where the
synchrotron self-absorption frequency is identified, provides three
independent equations for the synchrotron frequency, the synchrotron flux and the
black-body flux. A fourth equation is needed to fully constrain the system. Luckily, as it turns out, 
the electron and magnetic energy depend sensitively on $R$ in opposite
ways and the total energy is minimized at some radius, in which the
electrons and the magnetic field are roughly at equipartition. Thus the condition that the 
source energy is ``reasonable" provides a robust estimate of $R$. 
We denote this radius, where the energy is minimal as $R_{eq}$ and the
corresponding minimal energy as $E_{eq}$. Thus, a 
single measurement of synchrotron self-absorption frequency, $\nu_a$, and
flux, $F_{\nu,a}$, provides a robust, almost model independent, estimate of
the source size and its minimal energy. 

An extension to the relativistic case is important, because of the existence 
of synchrotron sources that involve relativistic bulk motion: 
jets in Gamma-Ray Bursts (GRB; e.g., Piran 2004),
Active Galactic Nuclei (AGN; e.g., Krolik 1998), 
relativistic Type Ibc supernovae (e.g., Soderberg et al. 2010b);
relativistic jets in tidal disruption event candidates (e.g., Zauderer et al.
2011) and others. Kumar \& Narayan (2009) derived the constraints that 
synchrotron emission can put on a relativistic source in the context of 
the prompt optical and gamma-ray observations of GRB 080319B  
(the `naked-eye burst'). This work was used later in the context of a tidal disruption event
candidate (Zauderer et al. 2011). 

Following the spirit of Kumar \& Narayan (2009), we present here
an explicit general extension of the equipartition
arguments, previously derived for Newtonian sources, to
sources that display relativistic bulk
motion. This generalization introduces a new free
parameter, the source's bulk Lorentz factor, $\Gamma$. The solution
requires an additional equation: the relation between $R$, $\Gamma$
and the time in the observer frame. Because of relativistic beaming,
geometrical effects\footnote{Note that since the true geometrical
parameters that affect the observations are the area and volume, the commonly 
used Newtonian formalism relies on the assumption of spherical symmetry,
without explicitly deriving the possible effects of deviations from that 
symmetry on the results.} could be important\footnote{We consider sources 
that move along (or close enough to) the line of sight, otherwise the
radiation will be beamed away from us.}. We consider, therefore, a general
source geometry. In particular we examine a wide jet with a half-opening angle
$\theta_j \gae 1/\Gamma$ and a narrow jet with $\theta_j < 1/\Gamma$. 

In order to make this paper easy to use and to aid the interested reader in
finding the relevant equations quickly, in \S 2 we give a full description of the system
and provide the formulae that enable to determine the radius
and minimal total energy of the system in terms of the observables and the
geometry. The detailed derivation of these formulae can be found in \S 3.
In \S 4 we consider the effects of different geometry and of additional
energetic components that do not contribute directly to the observed emission. 
In many cases, and in particular for nearby objects, the self-absorption
frequency is not identified but the radius of the source is directly
measured. We present the analysis of such systems in \S 5. Finally, in \S 6 we
examine the case when the synchrotron-self-Compton component is
observed and securely identified. In this case, minimization of the total
energy is not necessary and all the parameters of the system can be solved
unambiguously. We summarize our results and consider some astrophysical
implications in \S 7.

\section{Description of the system and summary of main results}

Consider a source that produces synchrotron emission. The source is located 
at a redshift $z$ with a luminosity distance $d_L$. It is characterized by 
an observed peak specific flux, $F_{\nu,p}$ at a frequency $\nu_p$. The
synchrotron emitting system is described by five physical quantities: 
The total number of electrons within the observed region, $N_e$, the 
volume averaged magnetic field strength perpendicular to the line of
  sight (in the source co-moving frame), $B$, the LF of the electrons 
that radiate at $\nu_p$, $\gamma_e$, the size of the emitting region, $R$, 
and the LF of the source, $\Gamma$. 

\begin{figure}
\begin{center}
\includegraphics[width=5cm, angle = 0]{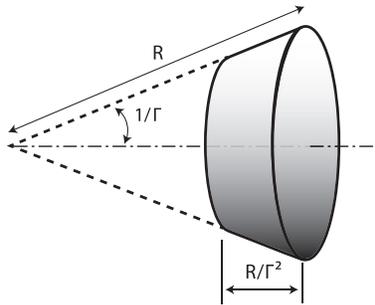}
\end{center}
\caption{We consider sources that move along or close enough to the line of
sight. Due to relativistic beaming, we can only detect emission within an
angle of $1/\Gamma$ with respect to the line of sight.  Therefore, unless the jet is very narrow with 
$\theta_j < 1/\Gamma$, the effective half-opening angle of the outflow
emitting region is 
$\approx 1/\Gamma$.  At a fixed observed time, the emission can only
be observed from a lab-frame width of order $R/\Gamma^2$, where the distance from the origin 
of the outflow is $R$. Thus, the region from which emission can be observed (shaded region of this
figure) has area $\pi R^2/\Gamma^2$ and volume $\pi R^3/\Gamma^4$. We denote
this region as the ``observed region''. For systems with a 
different geometry, the emitting region of the outflow, with area $A$ and volume $V$, can be
parameterized in terms of the fractions 
$f_A \equiv A/(\pi R^2/\Gamma^2)$ and 
$f_V \equiv V/(\pi R^3/\Gamma^4)$.}
\label{fig1} 
\end{figure}

In a relativistic outflow, at a fixed observed time $t$ from its onset, we can observe 
emission which comes mostly from a region within an angle of $1/\Gamma$ with
respect to the line of sight and from a lab-frame width of order $R/\Gamma^2$ (see Fig. \ref{fig1}).
We denote this region, from where emission can potentially be observed, as the ``observed region''.
Its area is $\pi R^2/\Gamma^2$ and its volume is $\pi R^3/\Gamma^4$.  
The source of the emission is not necessarily confined to the observed
region. Parts of the source that are outside of the observed region have no
effect on the observed emission (since photons generated outside of the
observed region cannot be observed anyway). In this case, the calculation
remains the same (that is, the factors $f_A$ and $f_V$, defined below, equal unity) and
the estimated energy must be multiplied by $\sim 2 \theta_j^2 \Gamma^2$ reflecting the
additional energy that is not observed directly (see \S \ref{Wide_section}).
However, if the source does not fill the entire observed region, the
calculation is affected. The effective source geometry is determined by the
total area, $A$, and volume, $V$, that are within the observed region. Thus,
it is convenient to parameterize the source geometry by the fractions of the 
observed region's area and volume that are filled by the source:
$f_A \equiv A/(\pi R^2/\Gamma^2) \leq 1$, and  
$f_V \equiv V/(\pi R^3/\Gamma^4) \leq 1$, which we denote as the area and 
the volume filling factors.  Note that in
the case of a continuous outflow, where the flow is wider 
than $\sim R/\Gamma^2$, this formalism applies only to the emission of a 
region (or ``blob'') that dominates the observed emission, whose width is 
$R/\Gamma^2$.
Note that the Newtonian equipartition solution usually assumes
a spherical source. In this case the volume filling factor   
$f_V$ equals $4/3$ and not unity (as one would have expected).

We assume a power-law electron energy distribution, which is characterized by
a minimal electron LF and a power-law index $p$, assumed to be $p > 2$ (the
exact value of $p$ will only be relevant when we consider the 
synchrotron-self-Compton case and the case when $\nu_m < \nu_a$ -- see below).  
Most of the electrons have energies around this minimal electron LF and
they emit at the synchrotron frequency $\nu_m$.  The synchrotron
self-absorption frequency, $\nu_a$, could be either above or below $\nu_m$.
For $\nu_m < \nu_a$, then the peak frequency is $\nu_p = \nu_a$, 
in which case $F_\nu \propto \nu^{5/2}$ for $\nu_m < \nu< \nu_a$ and 
$F_\nu \propto \nu^2$ for $\nu < \nu_m$. For $\nu_a < \nu_m$, then 
$\nu_p = \nu_m$, in which case $F_\nu \propto \nu^{1/3}$ for $\nu_a < \nu < \nu_m$ and
$F_\nu \propto \nu^{2}$ for $\nu < \nu_a$. Thus, $\nu_p = \max(\nu_a,\nu_m)$.
To take account of these two possibilities, if both $\nu_a$ and $\nu_m$ can be identified
in the spectrum, we define 
\begin{equation}
\eta \equiv \left\{ \begin{array}{ll} 
{\nu_m}/{\nu_a} & \textrm{if $\nu_a < \nu_m$} \\
1 & \textrm{if $\nu_a > \nu_m$}, \\
\end{array} \right.
\end{equation} 
or $\eta \equiv \nu_p/\nu_a$. This allows us to consider the most general
spectral shape. We assume that 
the observed peak frequency, $\nu_p$, is smaller than the cooling frequency, 
and we ignore the effect of electron cooling. 

In the rest of this section, we will present, without derivation, a summary
of the main equations describing relativistic equipartition. These include 
the estimates of the radius, Lorentz factor and minimal energy. The derivation 
of these equations, as well as other quantities, is presented in \S \ref{Main_section}.    

\begin{figure}
\begin{center}
\includegraphics[width=12cm, angle = 0]{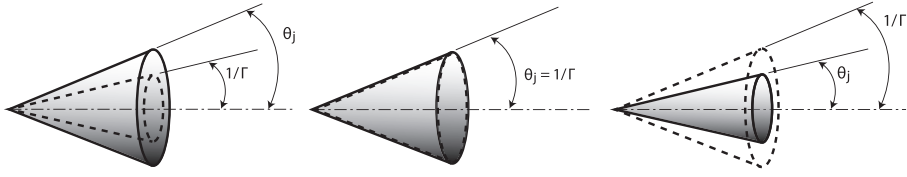}
\end{center}
\caption{Different types of relativistic outflows. From left to right:
A wide jet, where $\theta_j > 1/\Gamma$; a jet, where 
$\theta_j \approx 1/\Gamma$, and a narrow jet, where $\theta_j <
1/\Gamma$. With $f_A=f_V=1$ all equations in this paper correspond to a jet with 
$\theta_j \approx 1/\Gamma$; however, any general geometry can be considered
by using appropriate values for $f_A$ and $f_V$ (for the case of a 
narrow and a wide jet see \S \ref{Narrow_section} and 
\S \ref{Wide_section}).}
\label{fig2} 
\end{figure}

Energy minimization arguments, which result in a rough equipartition
between the electrons and the magnetic field, allow us to constrain four of the
five physical parameters of the system.  Therefore, we can express the
equipartition radius, $R_{eq}$, and minimal total energy, $E_{eq}$, as  
functions of the observables ($F_p$, $d_L$, $\nu_p$, $\eta$, $z$), the
geometrical parameters ($f_A$ and $f_V$) and one of the physical 
parameters (we choose the bulk LF of the source) as:
\begin{equation} 
R_{eq} \approx (1.7\times10^{17} {\rm cm}) \, \left[ F_{p,mJy}^{\frac{8}{17}} \, d_{L,28}^{\frac{16}{17}} \, \nu_{p,10}^{-1} \, \eta^{\frac{35}{51}} \,
(1+z)^{-\frac{25}{17}} \right] \, \frac{\Gamma^{\frac{10}{17}}}{f_A^{\frac{7}{17}} \, f_V^{\frac{1}{17}}},
\end{equation}
\begin{equation} 
E_{eq} \approx (2.5\times10^{49} {\rm erg}) \, \left[ F_{p,mJy}^{\frac{20}{17}} \, d_{L,28}^{\frac{40}{17}} \, \nu_{p,10}^{-1} \, \eta^{\frac{15}{17}} \,
(1+z)^{-\frac{37}{17}} \right] \,\frac{f_V^{\frac{6}{17}} }{f_A^{\frac{9}{17}} \, \Gamma^{\frac{26}{17}}}.
\end{equation}
Here, we have used $F_{p,mJy}=F_{\nu,p}/{\rm mJy}$ and, throughout the paper, we use the usual
notation $Q_n = Q/10^n$ in cgs units. For clarity, here and elsewhere, the observed quantities are
grouped and written between square brackets to distinguish them clearly from
the physical parameters of the system. The next step is to estimate the LF of
the source.  If it is related to the time since the onset of the
relativistic outflow as $t\approx R(1+z)/(2c\Gamma^2)$, then the radius, bulk
LF and minimal total energy are given by:
\begin{equation} 
R_{eq} \approx (7.5\times10^{17} {\rm cm}) \, \left[ F_{p,mJy}^{\frac{2}{3}} \, d_{L,28}^{\frac{4}{3}} \, \nu_{p,10}^{-\frac{17}{12}} \, \eta^{\frac{35}{36}} \,
(1+z)^{-\frac{5}{3}} \, t_d^{-\frac{5}{12}} \right] \, f_A^{-\frac{7}{12}} \, f_V^{-\frac{1}{12}},
\end{equation}
\begin{equation} 
\Gamma \approx 12 \, \left[ F_{p,mJy}^{\frac{1}{3}} \, d_{L,28}^{\frac{2}{3}} \, \nu_{p,10}^{-\frac{17}{24}} \, \eta^{\frac{35}{72}} \,
(1+z)^{-\frac{1}{3}} \, t_d^{-\frac{17}{24}} \right] \, f_A^{-\frac{7}{24}} \, f_V^{-\frac{1}{24}},
\end{equation}
\begin{equation} 
E_{eq} \approx (5.7\times10^{47} {\rm erg}) \, \left[ F_{p,mJy}^{\frac{2}{3}} \, d_{L,28}^{\frac{4}{3}} \, \nu_{p,10}^{\frac{1}{12}} \, \eta^{\frac{5}{36}} \,
(1+z)^{-\frac{5}{3}} \, t_d^{\frac{13}{12}} \right] \, f_A^{-\frac{1}{12}} \, f_V^{\frac{5}{12}},
\end{equation}
where the time, $t_d$, is measured in days. With $f_A=f_V=1$ these equations 
describe the energy within an outflow with a half-opening angle 
$\theta_j = 1/\Gamma$ (see Figs. \ref{fig1} and \ref{fig2}).  In \S \ref{Narrow_section} and 
\S \ref{Wide_section} we discuss the implications of a narrow jet 
($\theta_j < 1/\Gamma$) and a wider ($\theta_j > 1/\Gamma$) outflow. We also 
have not considered here the energy of the electrons that radiate at 
$\nu_m$ when $\nu_m < \nu_a$, which is discussed in \S \ref{nu_m_section}.
In addition, a similar analysis can be done for a source for which we know its
size but ignore the location of $\nu_a$, as discussed in \S \ref{Known_size}.

Alternatively, if a measurement of the synchrotron-self-Compton (SSC) component is
available (and securely identified), then one can abandon the energy minimization
argument (see \S 6).  In this case, we can express the radius of emission as a function
of $\Gamma$ as:
\begin{eqnarray} 
R &\approx& (1\times10^{17} {\rm cm}) \, [5(525)^{p-3}]^{\frac{1}{2(2+p)}} \, \Bigg[
F_{p,mJy}^{\frac{1}{2}} \, d_{L,28} \, \nu_{p,10}^{-\frac{3+2p}{2(2+p)}} \, \eta^{\frac{5(1+p)}{6(2+p)}}
\nonumber \\
&& \times (1+z)^{-\frac{5+3p}{2(2+p)}} 
\, \left(\frac{F_{\nu,p}}{F_{\nu}^{SSC}}\right)^{\frac{1}{2(2+p)}} \,
\left(\frac{\nu_p}{\nu_{obs}^{SSC}}\right)^{\frac{p-1}{4(2+p)}} \Bigg]
\, f_A^{-\frac{1+p}{2(2+p)}} \, \Gamma^{\frac{1+p}{2(2+p)}},
\end{eqnarray}
where $\nu_{obs}^{SSC}$ and $F_{\nu}^{SSC}$ are the measured frequency 
and specific flux of the SSC component, and the measured frequency is 
{\it above} the SSC peak.  In a similar way as done above, relating the bulk LF of the 
source to the time since the onset of the relativistic explosion allows 
us to determine $R$ (and all other physical parameters) only as 
function of observables.  Here, we show $R$ and $\Gamma$: 

\begin{eqnarray} 
R &\approx& (1\times10^{17} {\rm cm}) \, C_1^{\frac{2}{7+3p}} \, C_2^{\frac{2(1+p)}{7+3p}} \,
\Bigg[F_{p,mJy}^{\frac{2(2+p)}{7+3p}} \, d_{L,28}^{\frac{4(2+p)}{7+3p}} \, \nu_{p,10}^{-\frac{2(3+2p)}{7+3p}} \, \eta^{\frac{10(1+p)}{3(7+3p)}}
\nonumber \\
&& \times (1+z)^{-\frac{9+5p}{7+3p}} \, t_d^{-\frac{1+p}{7+3p}} \,
\left(\frac{F_{\nu,p}}{F_{\nu}^{SSC}}\right)^{\frac{2}{7+3p}} \,
\left(\frac{\nu_p}{\nu_{obs}^{SSC}}\right)^{\frac{p-1}{7+3p}} \Bigg]
\,f_A^{-\frac{2(1+p)}{7+3p}},  
\end{eqnarray}
\begin{eqnarray} 
\Gamma &\approx& C_1^{\frac{1}{7+3p}} \, C_2^{\frac{4(2+p)}{7+3p}} \,
\Bigg[F_{p,mJy}^{\frac{2+p}{7+3p}} \, d_{L,28}^{\frac{2(2+p)}{7+3p}} \, \nu_{p,10}^{-\frac{3+2p}{7+3p}} \, \eta^{\frac{5(1+p)}{3(7+3p)}}
\nonumber \\
&& \times (1+z)^{-\frac{1+p}{7+3p}} \, t_d^{-\frac{2(2+p)}{7+3p}} \,
\left(\frac{F_{\nu,p}}{F_{\nu}^{SSC}}\right)^{\frac{1}{7+3p}} \, 
\left(\frac{\nu_p}{\nu_{obs}^{SSC}}\right)^{\frac{p-1}{2(7+3p)}} \Bigg]
f_A^{-\frac{1+p}{7+3p}},  
\end{eqnarray}
where $C_1 \approx 5(525)^{p-3}$ and $C_2 \approx 4.4$, and the rest of the
parameters can be found in the Appendix, including the total energy in electrons
and magnetic field.

\section{Derivation of the radius estimate and the minimal total energy} \label{Main_section}

A synchrotron emitting system is characterized by three equations: 
the synchrotron frequency, the synchrotron flux and the black-body
flux. The observed synchrotron frequency is
\begin{equation} \label{nu_p}
\nu_p = \frac{e B \gamma_e^2 \Gamma}{2 \pi m_e c (1+z)},
\end{equation}
where $e$ is the electron charge, $m_e$ is the electron 
mass, $c$ is the speed of light and $z$ is the redshift.
The observed synchrotron maximum specific flux, at $\nu_p$,
  is\footnote{For the
  precise numerical prefactors of eqs. (\ref{nu_p}) and (\ref{f_p}), which
  depend (weakly) on $p$, see Wijers \& Galama (1999); here we have used 
  approximate values for $p \gae 2$.}  
\begin{equation} \label{f_p}
F_{\nu,p} = \frac{\sqrt{3} e^3 B N_e \Gamma^3 (1+z)}{\pi d_L^2 m_e c^2}, 
\end{equation}
where we have used the fact that the emission is beamed into a solid angle of $\pi/\Gamma^2$.
This expression is different than eq. (5) of Sari et al. (1998; see, also,
Kumar \& Narayan 2009) that uses $ N_{e,iso}$, the isotropic equivalent number
of electrons, rather than $N_e$, the number of electrons within the observed
region ($N_{e,iso} = 4 \Gamma^2 N_e$). This additional factor of 
$4$ introduces small corrections when taking the Newtonian limit ($\Gamma=1$):   
Eqs. (\ref{equi_radius}), (\ref{equi_energy}), (\ref{equi_radius2}),
(\ref{equi_energy2}), (\ref{equi_B}) and (\ref{equi_energy4}), should be multiplied by  $4^{1/17}$,
$4^{11/17}$, $4^{{1/(13+2p)}}$, $4^{{11/(13+2p)}}$, $4^{2/7}$ and
$4^{4/7}$, respectively.

The black-body specific flux, at frequency $\nu \le \nu_a$, is given by 
\begin{equation} \label{f_BB}
F_{\nu,BB} = 2 \nu^2 (1+z)^3 \Gamma m_e \gamma_e \frac{A}{d_L^2},
\end{equation}
where $A = f_A \pi R^2/\Gamma^2$ and we have used an equivalent
effective black-body temperature, $kT$, as the energy of the electrons radiating at the peak, 
$k T \approx m_e c^2 \gamma_e$. The flux at $\nu_a$ is\footnote{ 
The right-hand side of eq. (\ref{flux_at_nu_a}) should be multiplied by a numerical
factor that depends on the observed synchrotron spectrum above the peak (Shen
\& Zhang 2009). For simplicity, here we take this factor to 
be $\sim 3$, which is an approximate average value for a range of typical observed
synchrotron spectra.}:
\begin{equation} \label{flux_at_nu_a}
F_{\nu_a,BB} = F_{\nu,p} \eta^{-\frac{1}{3}}.
\end{equation}

Using eqs. (\ref{nu_p})--(\ref{flux_at_nu_a}) we can solve for three of the five
physical parameters, $\gamma_e$, $N_e$ and $B$,  as functions of the
observables ($F_p$, $d_L$, $\nu_p$, $\eta$, $z$), the remaining two 
physical parameters ($R$ and $\Gamma$), and the geometrical parameters ($f_A$ and $f_V$):
\begin{equation} \label{gamma_e}
\gamma_e = \frac{3 F_{\nu,p} d_L^2 \eta^{\frac{5}{3}} \Gamma}{2 \pi \nu_p^2 (1+z)^3 m_e
  f_A R^2 } \approx 525 \, \left[F_{p,mJy} \, d_{L,28}^2 \, \nu_{p,10}^{-2} \, 
\eta^{\frac{5}{3}} \, (1+z)^{-3}\right] \, \frac{\Gamma}{f_A \, R_{17}^{2}} ,
\end{equation}
\begin{equation} \label{N_e}
N_e = \frac{9 c F_{\nu,p}^3 d_L^6 \eta^{\frac{10}{3}}}{8 \sqrt{3} \pi^2 e^2 m_e^2 \nu_p^5 (1+z)^8 f_A^2 R^4}
\approx 1 \times 10^{54} \,  \left[ F_{p,mJy}^3 \, d_{L,28}^6 \, \nu_{p,10}^{-5} \,
\eta^{\frac{10}{3}} \,(1+z)^{-8} \right] \,\frac{1}{ f_A^{2} \, R_{17}^{4}},
\end{equation}
\begin{equation} \label{B}
B =  \frac{8 \pi^3 m_e^3 c \nu_p^5 (1+z)^7 f_A^2 R^4}{9 e F_{\nu,p}^2 d_L^4 \eta^{\frac{10}{3}} \Gamma^3}
\approx (1.3 \times 10^{-2} \, {\rm G}) \, \left[ F_{p,mJy}^{-2} \, d_{L,28}^{-4} \, \nu_{p,10}^5 \,
\eta^{-\frac{10}{3}} \, (1+z)^7 \right] \, \frac{f_A^{2} \, R_{17}^4}{\Gamma^{3}}.
\end{equation}

The  energy in electrons within the observed region is
\begin{eqnarray} \label{E_e}
E_e &=& N_e m_e c^2 \gamma_e \Gamma = \frac{27 c^3 F_{\nu,p}^4 d_L^8 \eta^5 \Gamma^2}{16 \sqrt{3} \pi^3 e^2
  m_e^2 \nu_p^7 (1+z)^{11} f_A^3 R^6} \nonumber \\
&\approx&  (4.4 \times 10^{50} {\rm erg}) \, \Big[ F_{p,mJy}^4 \, d_{L,28}^8 \, \nu_{p,10}^{-7} \, \eta^{5}
\, (1+z)^{-11} \Big] \, \frac{ \Gamma^2}{f_A^{3} \, R_{17}^{6}},
\end{eqnarray}
while the energy in the magnetic field is
\begin{eqnarray} \label{E_B}
E_B &=& \frac{(B \Gamma)^2}{8 \pi} V =  \frac{8 \pi^6 m_e^6 c^2 \nu_p^{10}
  (1+z)^{14} f_A^4 f_V R^{11}}{81 e^2 F_{\nu,p}^4 d_L^8 \eta^{\frac{20}{3}} \Gamma^8} \nonumber \\
&\approx&  (2.1 \times 10^{46} {\rm erg}) \, \left[  F_{p,mJy}^{-4} \, d_{L,28}^{-8} \, \nu_{p,10}^{10} \, \eta^{-\frac{20}{3}}
\, (1+z)^{14}  \right] \, \frac{f_A^{4} \, f_V \, R_{17}^{11}}{ \Gamma^{8}},
\end{eqnarray}
where $V = f_V \pi R^3 /\Gamma^4$. 

Eqs. (\ref{gamma_e})--(\ref{E_B}) are reduced to the Newtonian case for
$\Gamma=f_A=1$ (note, however, that additional factors of
powers of 4, mentioned following eq. (\ref{f_p}), should be added, and that 
$f_V=4/3$ in the {\it spherical} Newtonian case). In the Newtonian analysis $\eta=1$ is generally
used. According to the classical Newtonian equipartition argument, we minimize the total energy
to obtain the Newtonian equipartition radius and minimal total energy in terms
of the observables: 
\begin{eqnarray} 
R_N &\approx& (1.7\times10^{17} {\rm cm}) \,\left[ F_{p,mJy}^{\frac{8}{17}} \, d_{L,28}^{\frac{16}{17}} \, \nu_{p,10}^{-1} \, \eta^{\frac{35}{51}} \,
(1+z)^{-\frac{25}{17}}\right], \nonumber \\
E_N &\approx& (2.5\times10^{49} {\rm erg}) \, \left[ F_{p,mJy}^{\frac{20}{17}} \, d_{L,28}^{\frac{40}{17}} \, \nu_{p,10}^{-1} \, \eta^{\frac{15}{17}} \,
(1+z)^{-\frac{37}{17}}\right]. 
\end{eqnarray}
Generalizing to the relativistic non-spherical symmetric case we can express, now, the total energy using $R_N$ and $E_N$ as:
\begin{equation} \label{general_E}
E= E_e + E_B= E_N \left(\frac{f_V^{\frac{6}{17}}}{f_A^{\frac{9}{17}} \Gamma^{\frac{26}{17}}}\right)
\left[\frac{11}{17}\left(\frac{R}{R_{eq}}\right)^{-6} +
\frac{6}{17}\left(\frac{R}{R_{eq}}\right)^{11} \right].
\end{equation}
$R_{eq}$ (in the Newtonian case $R_{eq}=R_N$) is the relativistic equipartition radius:
\begin{equation} \label{equi_radius}
R_{eq} \equiv \, R_N \,\frac{\Gamma^{\frac{10}{17}}}{f_A^{\frac{7}{17}} \, f_V^{\frac{1}{17}}} 
\approx (1.7\times10^{17} {\rm cm}) \, \left[ F_{p,mJy}^{\frac{8}{17}} \, d_{L,28}^{\frac{16}{17}} \, \nu_{p,10}^{-1} \, \eta^{\frac{35}{51}} \,
(1+z)^{-\frac{25}{17}} \right] \, \frac{\Gamma^{\frac{10}{17}}}{f_A^{\frac{7}{17}} \, f_V^{\frac{1}{17}}}.
\end{equation}
The total energy is minimized with respect to $R$ at $R_{eq}$, with $E_B \approx (6/11) E_e$.
Since the total energy is a very strong function of radius,  $R_{eq}$ provides
a robust estimate of $R$, unless we allow the total energy to be significantly
higher than the minimal allowed total energy.

Examination of eq. (\ref{equi_radius}) reveals that $R_{eq}$ varies only
weakly with  variation in the geometry.  Specifically, it is insensitive 
to the volume filling factor, $f_V$, and it depends only weakly on 
the area filling factor $f_A$. A considerable deviation from spherical 
symmetry is required to affect the radius estimate. Moreover, $R_{eq}$ increases with 
$\Gamma$, thus the application of the Newtonian estimate to 
an ultrarelativistic source results in a significant underestimate of
its radius. 

In the relativistic case, the total energy, eq. (\ref{general_E}), depends on two
unknowns: $R$ and $\Gamma$. For any given $\Gamma$ the energy is minimized at 
$R=R_{eq}$. However, if we choose $R=R_{eq}(\Gamma)$ then the value in square 
brackets of eq. (\ref{general_E}) is just unity and $E \propto \Gamma^{-26/17}$
Hence there is no global minimum for this function and we must determine $\Gamma$
independently.   We need now another relation that will enable us to express 
$\Gamma$ as a function of $R$. To obtain this
relation we introduce an extra observable, $t$, the time, in the observer frame, since the
onset of the relativistic outflow. In most astrophysical scenarios $\Gamma$
evolves on a time scale comparable to, or longer than, $t$ and:
\begin{equation} \label{R-observed_time}
t \approx \frac{R (1-\beta) (1+z)}{\beta c},
\end{equation} 
where $\beta$ is the velocity of the outflow at observer time $t$. If the
time of the onset of the outflow is known then a single measurement of the synchrotron spectrum
is enough and eqs. (\ref{equi_radius}) and (\ref{R-observed_time}) are
solved simultaneously\footnote{Strictly speaking, in this case we have to substitute
  $\Gamma(R)$ from eq. (\ref{R-observed_time}) in eqs. (\ref{E_e}) and
  (\ref{E_B}) and minimize the total energy with respect to $R$. However, it
  can be shown that this procedure yields almost identical results to solving
  eqs. (\ref{equi_radius}) and (\ref{R-observed_time}) simultaneously.} to determine $R$ and $\Gamma$.
In the extreme relativistic limit $\Gamma \gg 1$, $t\approx
R(1+z)/(2c\Gamma^2)$ and we find a radius
\begin{equation} \label{equi_radius_ultra}
R_{eq} \approx (7.5\times10^{17} {\rm cm}) \, \left[ F_{p,mJy}^{\frac{2}{3}} \, d_{L,28}^{\frac{4}{3}} \, \nu_{p,10}^{-\frac{17}{12}} \, \eta^{\frac{35}{36}} \,
(1+z)^{-\frac{5}{3}} \, t_d^{-\frac{5}{12}} \right] \, f_A^{-\frac{7}{12}} \, f_V^{-\frac{1}{12}},
\end{equation}
and a bulk LF given by
\begin{equation} \label{equi_LF_ultra}
\Gamma \approx 12 \, \left[ F_{p,mJy}^{\frac{1}{3}} \, d_{L,28}^{\frac{2}{3}} \, \nu_{p,10}^{-\frac{17}{24}} \, \eta^{\frac{35}{72}} \,
(1+z)^{-\frac{1}{3}} \, t_d^{-\frac{17}{24}} \right] \, f_A^{-\frac{7}{24}} \, f_V^{-\frac{1}{24}},
\end{equation}
where $t_d$ is the time measured in days.

If the onset of the outflow is unknown, then we need at least two epochs, $t_1$ and $t_2$, at
which $F_{\nu,p}$ and $\nu_p$ (and $\nu_a$ if it is not the peak frequency)
are measured. If $\Gamma(t_1) \sim \Gamma(t_2)$, we solve eqs. (\ref{equi_radius})
and (\ref{R-observed_time}) for  $R(t_2)$ and $R(t_1)$ and $t_2$ and 
$t_1$.
However, $\Gamma$ may evolve on a time scale comparable to $t$. Therefore if $t_2 \gg t_1$ it is
possible that $\Gamma(t_1) \nsim \Gamma(t_2)$. This case is identified if the
above procedure results in $R(t_2) \gg R(t_1)$. Then $t_2-t_1 \sim t$ and
$R(t_2)-R(t_1) \sim R$, and we can approximate the solution at $t_2$  using $t_2
\approx t$. In this case the solution of $R(t_1)$ cannot be trusted. 

Substitution of $R_{eq}$ into eq. (\ref{general_E}) yields the absolute minimal
total energy of the system. This energy  accounts only for the
electrons that radiate at $\nu_p$ and for the corresponding magnetic field.
For both components we consider only the energy within the observed region:
\begin{equation} \label{equi_energy}
E_{eq} = E_N \, \frac{f_V^{\frac{6}{17}} }{f_A^{\frac{9}{17}} \, \Gamma^{\frac{26}{17}}}
\approx (2.5\times10^{49} {\rm erg}) \, \left[ F_{p,mJy}^{\frac{20}{17}} \, d_{L,28}^{\frac{40}{17}} \, \nu_{p,10}^{-1} \, \eta^{\frac{15}{17}} \,
(1+z)^{-\frac{37}{17}} \right] \,\frac{f_V^{\frac{6}{17}} }{f_A^{\frac{9}{17}} \, \Gamma^{\frac{26}{17}}}.
\end{equation}
This lower limit decreases with $\Gamma$, and, therefore, it is less stringent 
for relativistic sources. This is driven mostly by the increased beaming, and 
thus the reduced area and volume within an angle of $\sim 1/\Gamma$. In
the relativistic limit, $\Gamma \gg 1$, we can use eq. (\ref{equi_LF_ultra})
to obtain:
\begin{equation} \label{equi_energy_ultra}
E_{eq} \approx (5.7\times10^{47} {\rm erg}) \, \left[ F_{p,mJy}^{\frac{2}{3}} \, d_{L,28}^{\frac{4}{3}} \, \nu_{p,10}^{\frac{1}{12}} \, \eta^{\frac{5}{36}} \,
(1+z)^{-\frac{5}{3}} \, t_d^{\frac{13}{12}} \right] \, f_A^{-\frac{1}{12}} \, f_V^{\frac{5}{12}}.
\end{equation}

The radius $R_{eq}$ that was obtained by minimizing the energy and assuming equipartition is a robust
estimate even if the system is out of equipartition. We define the
microphysical parameters, $\epsilon_e$ and $\epsilon_B$, as the fractions of
the total energy in electrons and magnetic field, respectively.  The energy 
is minimal for $\epsilon_B/\epsilon_e \approx 6/11$. The ratio,
$\epsilon \equiv (\epsilon_B/\epsilon_e)/(6/11)$,  parameterizes
the deviation from equipartition. The radius is multiplied by $\epsilon^{1/17}$ for the
Newtonian case, and by a factor of $\epsilon^{1/12}$ (and 
consequently $\Gamma$ is multiplied by a factor of $\epsilon^{1/24}$)
in the relativistic ($\Gamma \gg 1$) case. While the emission radius depends extremely weakly
on $\epsilon$, the total energy is a strong function of $\epsilon$ and
deviations from equipartition increase significantly 
the overall energy budget.  The energy is larger than the minimal total energy
by 
$\approx (11/17)\epsilon^{-6/17} + (6/17)\epsilon^{11/17}$ in the
Newtonian case and by 
$\approx (11/17)\epsilon^{-5/12} + (6/17)\epsilon^{7/12}$  
if the system is relativistic.

\section{The minimal (equipartition) energy}
Eq. (\ref{equi_energy}) provides an absolute lower limit to the energy of the system. 
This expression includes the energy of the electrons emitting at $\nu_p$ and the corresponding
magnetic field. Both terms are calculated within the observed region of
half-opening angle of $\sim 1/\Gamma$. 
We examine several cases in which additional
energy is ``hidden" in the system and is not observed directly, but it 
influences, of course, the overall  energy budget. However, before doing so we
consider the effect of the geometrical factors on the system. 

\subsection{Geometrical effects}

The effect of deviation from spherical geometry is opposite for $f_A$ and
$f_V$, for both the Newtonian, $E_{eq} \propto f_A^{-9/17} f_V^{6/17}$,
and the relativistic, $E_{eq} \propto f_A^{-1/12} f_V^{5/12}$, cases; see
eqs. (\ref{equi_energy}) and (\ref{equi_energy_ultra}), respectively.  

\subsubsection{Narrow jets} \label{Narrow_section}

A particularly interesting geometric effect is the one in 
a relativistic narrow jet with half-opening angle $\theta_j$ that is 
smaller than $1/\Gamma$ (see Fig. \ref{fig2}). In this case we define 
$f_{\theta} \equiv (\theta_j \Gamma)^2$ and both geometric factors satisfy:
$f_A = f_V = f_{\theta}$. Substituting these values into eqs. (\ref{equi_radius_ultra}),
(\ref{equi_LF_ultra}) and (\ref{equi_energy_ultra}) we
find $R \propto f_{\theta}^{-2/3}$, $\Gamma \propto f_{\theta}^{-1/3}$ and $E_{eq}
\propto f_{\theta}^{1/3}$. Since $f_{\theta} < 1$, this implies that the
radius and bulk LF of a narrow jet will be larger than in the case with
$f_A=f_V=1$; however, the resulting minimal energy will be smaller. Specifically, 
these quantities scale with $\theta_j$ as $R \propto \theta_j^{-4/5}$, 
$\Gamma \propto \theta_j^{-2/5}$ and $E_{eq} \propto \theta_j^{2/5}$.
Thus, a jet narrower than $1/\Gamma$ requires lower energy to produce the observed 
emission (although the decrease in energy is small given the weak
$\theta_j$-dependence of the minimal energy). 
The reason for this effect is not trivial (as there are
competing effects), but the main driver is the reduction in the area, which 
reduces $B$ and leads to a significant increase of $R_{eq}$
and $\Gamma$.  This results in a lower $E_{eq}$ than in the $f_A = f_V =1$ 
case, see eq. (\ref{equi_energy}).

\subsubsection{Wide outflows} \label{Wide_section}

The outflow's half-opening angle could be larger than $1/\Gamma$ (see Fig. \ref{fig2}). In
this case the overall energy of the source is larger, as additional energy
at the region $\theta_j > 1/\Gamma$ has negligible contribution to 
the observed emission .  The flow will carry an energy 
larger than the one calculated in eq. (\ref{equi_energy}), with $f_A=f_V=1$, 
by a factor of $4 \Gamma^2 (1 - \cos \theta_j)$. 
The ``true'' energy can be determined only if an independent
estimate of the jet opening angle is available (such as in GRBs, when a 
``jet break'' takes place and $\theta_j$ can be estimated, e.g., 
Sari et al. 1999). 

\medskip

\subsection{Unaccounted-for energy}

\subsubsection{Electrons that radiate at $\nu_m$} \label{nu_m_section}

Above we considered only the electrons that radiate at $\nu_p$. These
electrons are likely to carry most of the relativistic electron energy if
$\nu_p=\nu_m$. However, if $\nu_m < \nu_a$ most of the electrons' energy is
carried by the electrons with the minimal Lorentz factor, $\gamma_m$ (and
whose emission is self-absorbed). In this case the electrons' energy will be
larger than that of eq. (\ref{E_e}) (with $\eta=1$), by a factor of 
$(\gamma_m/\gamma_e)^{2-p}$,
where $p$ is the electron energy distribution power-law and $p>2$. In rare
cases $\nu_m$ can be identified in the spectrum. This can be done if
spectra at different epochs are available and one observes a
transition in the spectrum from
$F_\nu \propto \nu^2$ to $F_\nu \propto \nu^{5/2}$. In these cases
$(\gamma_m/\gamma_e)^{2-p} =
(\nu_m/\nu_a)^{(2-p)/2}$, so the radius estimate
is hardly modified, since it is only multiplied by
$(\nu_m/\nu_a)^{(2-p)/34}$ in eq. (\ref{equi_radius}) (with
$\eta =1$).  The total minimal energy is somewhat increased, since it is
multiplied by $(\nu_m/\nu_a)^{11(2-p)/34}$ in
eq. (\ref{equi_energy}) (with $\eta=1$). In the most common case
where $\nu_m$ is not measured it must be evaluated theoretically. This can be 
done if the electrons are known to be accelerated by a shock with LF similar 
to that of the source, $\Gamma$. In that case 
$\gamma_m = \chi_e (\Gamma-1)$ 
where $\chi_e = \frac{p-2}{p-1}\epsilon_e \frac{m_p}{m_e}$ and 
$\epsilon_e$ is the fraction of the protons energy that
goes into electrons and $m_p$ is the proton mass (if $\gamma_m$ is found to be
$\gamma_m < 2$, then one should use $\gamma_m = 2$).
With this, and following the same procedure as above of setting
$E_B \approx (6/11) E_e$, the radius where the energy is minimal becomes
\begin{eqnarray} \label{equi_radius2}
R_{eq} &\approx& (1\times10^{17} {\rm cm}) \, [21.8(525)^{p-1}]^{\frac{1}{13+2p}} \,
\chi_e^{\frac{2-p}{13+2p}} \, \Big[ F_{p,mJy}^{\frac{6+p}{13+2p}} \, d_{L,28}^{\frac{2(p+6)}{13+2p}}
\nonumber \\
&& \times \, \nu_{p,10}^{-1} \, (1+z)^{-\frac{19+3p}{13+2p}} \Big] \, 
f_A^{-\frac{5+p}{13+2p}} \, f_V^{-\frac{1}{13+2p}} \,
\Gamma^{\frac{p+8}{13+2p}} \, (\Gamma - 1)^{\frac{2-p}{13+2p}}.
\end{eqnarray}
The corresponding minimal total energy within the observed region is:
\begin{eqnarray} \label{equi_energy2}
E_{eq} &\approx& (1.3\times10^{48} {\rm erg}) \, [21.8]^{-\frac{2(p+1)}{13+2p}} \, 
[(525)^{p-1}\chi_e^{2-p}]^{\frac{11}{13+2p}} \, \Big[
F_{p,mJy}^{\frac{14+3p}{13+2p}} \, d_{L,28}^{\frac{2(3p+14)}{13+2p}} \nonumber \\
&& \times \, \nu_{p,10}^{-1} \, (1+z)^{-\frac{27+5p}{13+2p}} \Big]  \,
f_A^{-\frac{3(p+1)}{13+2p}} \, f_V^{\frac{2(p+1)}{13+2p}} \, 
\Gamma^{-\frac{5p+16}{13+2p}} \, (\Gamma - 1)^{-\frac{11(p-2)}{13+2p}}.
\end{eqnarray}
The last two expressions reduce to eqs. (\ref{equi_radius}) and
(\ref{equi_energy}) with $\eta=1$ for $p=2$ (when all electrons 
carry a similar amount of energy). For the Newtonian case, 
$\gamma_m = 2$ and $p=3$, one obtains the
solution found by Chevalier (1998).

\subsubsection{Hot protons} 

If the source contains protons it is reasonable to expect that these  
take a significant share of the
total internal and bulk energy. For example, observations indicate that in shock
heated gas (for example, in GRB afterglows, see, e.g., Panaitescu \& Kumar 2002) 
most of the energy is carried by hot protons. The exact
fraction of the total energy carried by other components is unknown, but these
observations suggest that the fraction carried by electrons, $\epsilon_e$, is
typically $\sim 0.1$ in relativistic shocks and lower in Newtonian shocks. Using this
parameterization, the energy carried by the hot protons is $E_p \approx
E_e/\epsilon_e$. This implies a total matter energy of $E_e + E_p = \xi E_e$, where
$\xi \equiv 1 + \epsilon_e^{-1}$. Similarly, the parameters at which the
energy is minimal are found by setting $E_B \approx (6/11) \xi E_e$. The
radius estimate is hardly modified, since it is only multiplied by
$\xi^{1/17}$, $\xi^{1/12}$ and $\xi^{1/(13+2p)}$ in eqs.
(\ref{equi_radius}), (\ref{equi_radius_ultra}) and (\ref{equi_radius2}),
respectively. The total minimal energy is somewhat increased, since it is multiplied
by $\xi^{11/17}$, $\xi^{7/12}$ and $\xi^{11/(13+2p)}$ in eqs.
(\ref{equi_energy}), (\ref{equi_energy_ultra}) and (\ref{equi_energy2}),
respectively.

\section{Systems with measured $R$ but unknown self-absorption frequency} \label{Known_size}.

There are cases, especially for Galactic and local universe sources, in which
we can resolve and measure the source's size on the sky and
determine $R \psi = \theta_{obs} d_A$, where 
$\psi \equiv \min(1/\Gamma,\theta_j)$,
$\theta_{obs}$ is the half-angular extent of the
source and $d_A = d_L (1+z)^{-2}$ is the angular distance.
However, for these sources we do not always have a measurement of $\nu_a$. We
can still estimate a minimal total energy carried by the magnetic field and by electrons that
radiate at the observed frequency $\nu$ at a flux $F_{\nu}$. This was
first done in the Newtonian case by Burbidge (1959; see, e.g., Nakar et al.
2005, for a recent example) and the relativistic
case, without considering any geometrical factors, 
was discussed in Dermer \& Atoyan (2004; see also Dermer \& Menon 2009). 
Determining the LF of electrons radiating at $\nu$ 
with (\ref{nu_p}) and the number of radiating electrons
within $1/\Gamma$ with (\ref{f_p}), we can determine the total energy of the
system. It is minimized once $E_B \approx (3/4) E_e$, which yields an equipartition 
magnetic field (see, e.g., Dermer \& Menon 2009)
\begin{equation} \label{equi_B}
B_{eq} \approx (5\times10^{-3} {\rm G}) \, \left[ F_{\nu,mJy}^{\frac{2}{7}} \,
\left(\frac{d_L}{10 {\rm kpc}}\right)^{-\frac{2}{7}} \, 
\nu_{10}^{\frac{1}{7}} \, (1+z)^{\frac{11}{7}} \, 
\left(\frac{\theta_{obs}}{10 {\rm mas}}\right)^{-\frac{6}{7}} \right] \,
f_V^{-\frac{2}{7}} \, \psi^{\frac{6}{7}} \, \Gamma^{-\frac{1}{7}}.
\end{equation}
The energy in the magnetic field can be determined with $B_{eq}$, and the total minimal energy
within the observed region, which is given by $E_e + E_B = (7/3) E_B$, is
\begin{equation} \label{equi_energy4}
E_{eq} \approx (2.8\times10^{40} {\rm erg}) \, \left[ F_{\nu,mJy}^{\frac{4}{7}} \,
\left(\frac{d_L}{10 {\rm kpc}}\right)^{\frac{17}{7}} \, 
\nu_{10}^{\frac{2}{7}} \, (1+z)^{-\frac{20}{7}} \, 
\left(\frac{\theta_{obs}}{10 {\rm mas}}\right)^{\frac{9}{7}} \right] \,
f_V^{\frac{3}{7}} \, \psi^{-\frac{9}{7}} \, \Gamma^{-\frac{16}{7}},
\end{equation}
where $F_{\nu,mJy} = F_{\nu}/mJy$. For these nearby sources we usually have 
the time of the onset of the outflow and can estimate $\Gamma$. This allows us to 
use (\ref{equi_energy4}) to estimate the absolute minimum total energy.

\section{Synchrotron-self-Compton emission} \label{SSC}

If synchrotron-self-Compton (SSC) emission is also observed, then there is no need to
minimize the total energy. This introduces two additional observables that
allow us to determine all parameters of the system without the need of
minimizing the total energy.  This was done 
by Chevalier \& Fransson (2006) and later by Katz (2012) for the Newtonian 
case. Here, we extend these estimates to the relativistic case, 
again, following the spirit of Kumar \& Narayan (2009; see also Dermer
\& Atoyan 2004). If the synchrotron emission peaks in the radio band
and the SSC observed emission is in the X-rays, then it is safe to assume, as
we do in the following, that the Klein-Nishina effects can be neglected.  

The SSC peak frequency, $\nu_p^{SSC}$, and the ratio of the synchrotron
to the SSC luminosities are
\begin{equation} \label{Compton-Y1}
\nu_p^{SSC} \approx \nu_p \gamma_e^2 ,
\end{equation}
and 
\begin{equation} \label{Compton-Y2}
\frac{B^2/8\pi}{U_{ph}} \approx \frac{\nu_p F_{\nu,p}}{\nu_p^{SSC} F_{\nu,p}^{SSC}},
\end{equation}
where $U_{ph}$ is the photon energy density (in the co-moving frame) 
and $F_{\nu,p}^{SSC}$ is the SSC peak flux. Note that eq. (\ref{Compton-Y1}) is
correct for both $\nu_a < \nu_m$ or $\nu_m < \nu_a$, since $\gamma_e$ corresponds
to the electrons radiating at $\nu_p = \max(\nu_a,\nu_m)$. The photon energy density can be
approximated as $U_{ph}=\frac{\nu_p F_{\nu,p}}{\Gamma^2 c}\frac{d_L^2}{R^2}$.

Consider an observed SSC frequency $\nu_{obs}^{SSC}$, such that $\nu_p^{SSC} < \nu_{obs}^{SSC}$,
with observed flux $F_{\nu}^{SSC}$.  These observations are related to the peak of
the SSC component as 
\begin{equation} \label{SSC_flux}
F_{\nu}^{SSC} = F_{\nu,p}^{SSC} \left(\frac{\nu_{obs}^{SSC}}{\nu_p^{SSC}} \right)^{-\frac{p-1}{2}}.
\end{equation}
Using eqs. (\ref{gamma_e}), (\ref{B}), and (\ref{Compton-Y1})--(\ref{SSC_flux}), 
we can solve for the radius of emission
\begin{eqnarray} \label{equi_radius4}
R &\approx& (1\times10^{17} {\rm cm}) \, [5(525)^{p-3}]^{\frac{1}{2(2+p)}} \, \Bigg[
F_{p,mJy}^{\frac{1}{2}} \, d_{L,28} \, \nu_{p,10}^{-\frac{3+2p}{2(2+p)}} \, \eta^{\frac{5(1+p)}{6(2+p)}}
\nonumber \\
&& \times (1+z)^{-\frac{5+3p}{2(2+p)}} 
\, \left(\frac{F_{\nu,p}}{F_{\nu}^{SSC}}\right)^{\frac{1}{2(2+p)}} \,
\left(\frac{\nu_p}{\nu_{obs}^{SSC}}\right)^{\frac{p-1}{4(2+p)}} \Bigg]
\, f_A^{-\frac{1+p}{2(2+p)}} \, \Gamma^{\frac{1+p}{2(2+p)}}.
\end{eqnarray}
This expression and eq. (\ref{R-observed_time}) allow us to
determine the radius of emission and $\Gamma$ of
the source.  We can then substitute the obtained values for $R$ and
$\Gamma$ in eqs. (\ref{gamma_e})-(\ref{E_B}) and obtain all physical
parameters of the emitting region. 
In the extreme relativistic limit $\Gamma \gg 1$, we can solve for all
these parameters analytically (see the Appendix for these expressions).

Finally, we note that for $\Gamma=1$ and $p=3$, eq. (\ref{equi_radius4}) 
reduces to the radius estimate in Katz (2012) for the Newtonian case, within a
factor of $\sim 2$. This small discrepancy appears simply because our
expression for the synchrotron frequency, eq. (\ref{nu_p}), is larger than the
one used by Katz (2012) by this same factor. 

\section{Summary}

We have extended the equipartition arguments of Newtonian synchrotron
sources in spherical geometry to include relativistic sources in general
geometry. This enables to derive robust estimates of the radius and of the
minimal total energy of the emitting region of a large variety of synchrotron
transient sources. It also enables to quantify the effect of the, typically
unknown, geometry on the robustness of these estimates.

We find that in the relativistic case the estimate of the emission radius is
increased by a factor of $\Gamma^{10/17}$ compared with the Newtonian case.
The lower limit on the energy (within a region of $\sim 1/\Gamma$) is lower by
$\Gamma^{-26/17}$ compared with the Newtonian one. Therefore, using the
Newtonian formalism for a relativistic source underestimates (overestimates)
the emission radius (lower limit on the energy). We show that in order to find
if relativistic corrections are needed, and to estimate $\Gamma$, at least two 
epochs of measurements are needed, or alternatively the time since the
onset of the outflow should be known.

The collimation of relativistic sources affects the energy lower limit.
Throughout the paper we considered an observed region of $\sim 1/\Gamma$;
however, considering a source with half-opening angle smaller (larger) than 
$1/\Gamma$ yields smaller (higher) lower limits.
A wider jet involves additional energy that we do not observe directly as it is beamed elsewhere, 
while the reason why a narrower jet requires lower energy is 
less trivial and is discussed above. 

The energy estimates discussed above involve the minimal energy 
(of the electrons and the magnetic field) required to produce the observed 
radiation. However, additional components in which energy is ``hidden" may exist in the system. 
These include: 1. The extra energy carried by electrons with minimal Lorentz
Factor $\gamma_m$ and whose synchrotron frequency $\nu_m$ is self-absorbed,
such that $\nu_m < \nu_a$, and 2. the energy carried by protons, if they are
present in the source. We consider their possible effect on the
total energy required. We find that these extra sources of energy hardly
change the emission radius, while the total minimal energy is increased.

Finally, we extend the Newtonian equipartition formalism to relativistic
sources in two other scenarios. First, for nearby sources, where we are able to identify the
angular size of the source on the sky, but the self-absorption frequency is
not identified. Second, for when a synchrotron-self-Compton component is
identified, in addition to the synchrotron self-absorption, and there are two
additional observables that enable us to directly determine all parameters of
the emitting region. Overall we find that relativistic corrections can be
important and that using the Newtonian formula for a relativistic source would
lead to significantly inaccurate results.

\acknowledgements{
RBD thanks Paz Beniamini for useful discussions. We thank Jessa Barniol for her help with
Figures 1 and 2. This work is supported by an Advanced ERC grant: GRB (RBD and TP), and by an ERC
starting grant and ISF grant no. 174/08 (EN).}

\section*{Appendix} 
\renewcommand{\theequation}{A\arabic{equation}}  
\setcounter{equation}{0}  
If a reliable measurement of the SSC flux is available, then there is no need
to minimize the total energy; all parameters of the emitting region can be
uniquely determined (see \S \ref{SSC}). In the extreme
relativistic limit $\Gamma \gg 1$, eq. (\ref{R-observed_time}) is 
$t \approx R(1+z)/(2c\Gamma^2)$, and we can solve for all
these parameters analytically as follows.  The radius of emission will be
given by eq. (\ref{equi_radius4}) as
\begin{eqnarray} 
R &\approx& (1\times10^{17} {\rm cm}) \, C_1^{\frac{2}{7+3p}} \, C_2^{\frac{2(1+p)}{7+3p}} \,
\Bigg[F_{p,mJy}^{\frac{2(2+p)}{7+3p}} \, d_{L,28}^{\frac{4(2+p)}{7+3p}} \, \nu_{p,10}^{-\frac{2(3+2p)}{7+3p}} \, \eta^{\frac{10(1+p)}{3(7+3p)}}
\nonumber \\
&& \times (1+z)^{-\frac{9+5p}{7+3p}} \, t_d^{-\frac{1+p}{7+3p}} \,
\left(\frac{F_{\nu,p}}{F_{\nu}^{SSC}}\right)^{\frac{2}{7+3p}} \,
\left(\frac{\nu_p}{\nu_{obs}^{SSC}}\right)^{\frac{p-1}{7+3p}} \Bigg]
\,f_A^{-\frac{2(1+p)}{7+3p}},  
\end{eqnarray}
and $\Gamma$ will be given by 
\begin{eqnarray} 
\Gamma &\approx& C_1^{\frac{1}{7+3p}} \, C_2^{\frac{4(2+p)}{7+3p}} \,
\Bigg[F_{p,mJy}^{\frac{2+p}{7+3p}} \, d_{L,28}^{\frac{2(2+p)}{7+3p}} \, \nu_{p,10}^{-\frac{3+2p}{7+3p}} \, \eta^{\frac{5(1+p)}{3(7+3p)}}
\nonumber \\
&& \times (1+z)^{-\frac{1+p}{7+3p}} \, t_d^{-\frac{2(2+p)}{7+3p}} \,
\left(\frac{F_{\nu,p}}{F_{\nu}^{SSC}}\right)^{\frac{1}{7+3p}} \, 
\left(\frac{\nu_p}{\nu_{obs}^{SSC}}\right)^{\frac{p-1}{2(7+3p)}} \Bigg]
f_A^{-\frac{1+p}{7+3p}},  
\end{eqnarray}
where $C_1 \approx 5(525)^{p-3}$ and $C_2 \approx 4.4$.  With these two
expressions, the rest of the parameters can be determined by
substituting them in eqs. (\ref{gamma_e})-(\ref{E_B}) as follows:
\begin{eqnarray} 
\gamma_e &\approx& 525 C_1^{-\frac{3}{7+3p}} \, C_2^{\frac{4}{7+3p}} \,
\Bigg[F_{p,mJy}^{\frac{1}{7+3p}} \, d_{L,28}^{\frac{2}{7+3p}} \, \nu_{p,10}^{-\frac{5}{7+3p}} \, \eta^{\frac{20}{3(7+3p)}}
\nonumber \\
&& \times (1+z)^{-\frac{4}{7+3p}} \, t_d^{-\frac{2}{7+3p}} \,
\left(\frac{F_{\nu,p}}{F_{\nu}^{SSC}}\right)^{-\frac{3}{7+3p}} \, 
\left(\frac{\nu_p}{\nu_{obs}^{SSC}}\right)^{\frac{3(1-p)}{2(7+3p)}} \Bigg]
f_A^{-\frac{4}{7+3p}},  
\end{eqnarray}
\begin{eqnarray} 
N_e &\approx& 1 \times 10^{54} \, C_1^{-\frac{8}{7+3p}} \, C_2^{-\frac{8(1+p)}{7+3p}} \,
\Bigg[F_{p,mJy}^{\frac{5+p}{7+3p}} \, d_{L,28}^{\frac{2(5+p)}{7+3p}} \, \nu_{p,10}^{-\frac{11-p}{7+3p}} \, \eta^{\frac{10(3-p)}{3(7+3p)}}
\nonumber \\
&& \times (1+z)^{-\frac{4(5+p)}{7+3p}} \, t_d^{\frac{4(1+p)}{7+3p}} \,
\left(\frac{F_{\nu,p}}{F_{\nu}^{SSC}}\right)^{-\frac{8}{7+3p}} \, 
\left(\frac{\nu_p}{\nu_{obs}^{SSC}}\right)^{\frac{4(1-p)}{7+3p}} \Bigg]
f_A^{-\frac{2(3-p)}{7+3p}},  
\end{eqnarray}
\begin{eqnarray} 
B &\approx& (1.3 \times 10^{-2} \, {\rm G}) \, C_1^{\frac{5}{7+3p}} \, C_2^{-\frac{4(4+p)}{7+3p}} \,
\Bigg[F_{p,mJy}^{-\frac{4+p}{7+3p}} \, d_{L,28}^{-\frac{2(4+p)}{7+3p}} \, \nu_{p,10}^{\frac{5(4+p)}{7+3p}} \, \eta^{-\frac{5(9+p)}{3(7+3p)}}
\nonumber \\
&& \times (1+z)^{\frac{4(4+p)}{7+3p}} \, t_d^{\frac{2(4+p)}{7+3p}} \,
\left(\frac{F_{\nu,p}}{F_{\nu}^{SSC}}\right)^{\frac{5}{7+3p}} \,
\left(\frac{\nu_p}{\nu_{obs}^{SSC}}\right)^{-\frac{5(1-p)}{2(7+3p)}} \Bigg]
f_A^{\frac{9+p}{7+3p}},
\end{eqnarray}
\begin{eqnarray} 
E_e &\approx& (4.4 \times 10^{50} \, {\rm erg}) \, C_1^{-\frac{10}{7+3p}} \, C_2^{\frac{4(1-p)}{7+3p}} \,
\Bigg[F_{p,mJy}^{\frac{2(4+p)}{7+3p}} \, d_{L,28}^{\frac{4(4+p)}{7+3p}} \, \nu_{p,10}^{-\frac{19+p}{7+3p}} \, \eta^{\frac{5(11-p)}{3(7+3p)}}
\nonumber \\
&& \times (1+z)^{-\frac{5(5+p)}{7+3p}} \, t_d^{-\frac{2(1-p)}{7+3p}} \,
\left(\frac{F_{\nu,p}}{F_{\nu}^{SSC}}\right)^{-\frac{10}{7+3p}} \,
\left(\frac{\nu_p}{\nu_{obs}^{SSC}}\right)^{\frac{5(1-p)}{7+3p}} \Bigg]
f_A^{-\frac{11-p}{7+3p}},
\end{eqnarray}
\begin{eqnarray} 
E_B &\approx& (2.1 \times 10^{46} \, {\rm erg}) \, C_1^{\frac{14}{7+3p}} \, C_2^{-\frac{2(21+5p)}{7+3p}} \,
\Bigg[F_{p,mJy}^{\frac{2p}{7+3p}} \, d_{L,28}^{\frac{4p}{7+3p}} \, \nu_{p,10}^{\frac{2(14+p)}{7+3p}} \, \eta^{-\frac{10(7-p)}{3(7+3p)}}
\nonumber \\
&& \times (1+z)^{\frac{7-5p}{7+3p}} \, t_d^{\frac{21+5p}{7+3p}} \, 
\left(\frac{F_{\nu,p}}{F_{\nu}^{SSC}}\right)^{\frac{14}{7+3p}} \,
\left(\frac{\nu_p}{\nu_{obs}^{SSC}}\right)^{-\frac{7(1-p)}{7+3p}} \Bigg]
f_A^{\frac{2(7-p)}{7+3p}} \, f_V.
\end{eqnarray}


\end{document}